\documentclass[aps,prb,twocolumn,superscriptaddress,showpacs,amsmath]{revtex4-1}

\usepackage{epsfig}

\begin{document}


\title{Extending spin coherence times of diamond qubits by high temperature annealing}


\author{T. Yamamoto}
\email[]{Current affiliation: National Institute for Materials Science, 1-1 Namiki, Tsukuba, Ibaraki 305-0044, Japan. email: YAMAMOTO.Takashi@nims.go.jp}
\affiliation{Japan Atomic Energy Agency, 1233 Watanuki, Takasaki, Gunma 370-1292, Japan}
\author{T. Umeda}
\affiliation{Institute of Applied Physics, University of Tsukuba, 1-1-1 Tennodai, Tsukuba, Ibaraki 305-8573 Japan}
\author{K. Watanabe}
\affiliation{National Institute for Materials Science, 1-1 Namiki, Tsukuba, Ibaraki 305-0044, Japan}
\author{S. Onoda}
\affiliation{Japan Atomic Energy Agency, 1233 Watanuki, Takasaki, Gunma 370-1292, Japan}
\author{M. L. Markham}
\affiliation{Element Six Limited, KingÕs Ride Park, Ascot, Berkshire, SL5 8BP, United Kingdom}
\author{D. J. Twitchen}
\affiliation{Element Six Limited, KingÕs Ride Park, Ascot, Berkshire, SL5 8BP, United Kingdom}
\author{B. Naydenov}
\affiliation{Institute for Quantum Optics, University of Ulm, D-89081, Ulm, Germany}
\author{L. P. McGuinness}
\affiliation{Institute for Quantum Optics, University of Ulm, D-89081, Ulm, Germany}
\author{T. Teraji}
\affiliation{National Institute for Materials Science, 1-1 Namiki, Tsukuba, Ibaraki 305-0044, Japan}
\author{S. Koizumi}
\affiliation{National Institute for Materials Science, 1-1 Namiki, Tsukuba, Ibaraki 305-0044, Japan}
\author{F. Dolde}
\affiliation{3rd Physics Institute and Research Center SCoPE, University of Stuttgart, D-70174, Stuttgart, Germany}
\author{H. Fedder}
\affiliation{3rd Physics Institute and Research Center SCoPE, University of Stuttgart, D-70174, Stuttgart, Germany}
\author{J. Honert}
\affiliation{3rd Physics Institute and Research Center SCoPE, University of Stuttgart, D-70174, Stuttgart, Germany}
\author{J. Wrachtrup}
\affiliation{3rd Physics Institute and Research Center SCoPE, University of Stuttgart, D-70174, Stuttgart, Germany}
\author{T. Ohshima}
\affiliation{Japan Atomic Energy Agency, 1233 Watanuki, Takasaki, Gunma 370-1292, Japan}
\author{F. Jelezko}
\affiliation{Institute for Quantum Optics, University of Ulm, D-89081, Ulm, Germany}
\author{J. Isoya}
\affiliation{Research Center for Knowledge Communities, University of Tsukuba, 1-2 Kasuga, Tsukuba, Ibaraki 305-8550, Japan}




\begin{abstract}
Spins of negatively charged nitrogen-vacancy (NV$^-$) defects in diamond are among the most promising candidates for solid-state qubits. The fabrication of quantum devices containing these spin-carrying defects requires position-controlled introduction of NV$^-$ defects having excellent properties such as spectral stability, long spin coherence time, and stable negative charge state. Nitrogen ion implantation and annealing enable the positioning of NV$^-$ spin qubits with high precision, but to date, the coherence times of qubits produced this way are short, presumably because of the presence of residual radiation damage. In the present work, we demonstrate that a high temperature annealing at 1000$^\circ$C allows 2 millisecond coherence times to be achieved at room temperature. These results were obtained for implantation-produced NV$^-$ defects in a high-purity, 99.99\% $^{12}$C enriched single crystal chemical vapor deposited diamond. We discuss these remarkably long coherence times in the context of the thermal behavior of residual defect spins. [Published in Physical Review B {\bf{88}}, 075206 (2013)]
\end{abstract}


\maketitle

\section{Introduction}
Quantum physics offers an enlightening route, capitalizing on the quantum nature of superposition and entanglement, to assist classical components in solving certain computational problems.~\cite{Nielsen} The physical implementation of a quantum register is however challenging since the quantum regime requires a high degree of isolation from the environment: otherwise, decoherence leads to loss of quantum information. Solid-state quantum bits (qubits) such as superconducting circuits,~\cite{Nakamura} quantum dots,~\cite{Hanson} and phosphorus donors in silicon,~\cite{Kane} which are promising for scalability, require low temperatures for operation. In contrast, electron spins of photoactive defects in diamond have shown significant potential as solid-state spin qubits operating under ambient conditions.~\cite{Jelezko3} The negatively charged nitrogen-vacancy (NV$^-$) defect, comprising a substitutional nitrogen with a vacancy at an adjacent lattice site, is a bright fluorescent center with a zero phonon line at 637~nm.~\cite{Davies1} The $|{\it{m}}_{\rm{s}}=0\rangle$ and $|{\it{m}}_{\rm{s}}=\pm1\rangle$ sublevels of ground-state spin triplet ($S=1$) are separated by $\sim2.88$~GHz in a zero magnetic field.~\cite{Reddy} The individual electron spins can be initialized optically, manipulated by microwave pulses, and then readout optically, all under ambient  conditions.~\cite{Jelezko,Jelezko2} In addition to their excellent properties as solid-state spin qubits, the spin coherence times $(T_2)$ exceed a millisecond at room temperature (RT).~\cite{Balasubramanian,Ishikawa,Jahnke} So far there has been progress for a scalable architecture comprised of two NV$^-$ qubits with magnetic coupling~\cite{Neumann2} and entanglement~\cite{Dolde} demonstrated for NV$^-$ centers with nanoscale separations introduced by nitrogen ion implantation. In this way, the technical approach of implantation appears promising for building a quantum register, but the short $T_2$ of implanted NV$^-$ is a considerable barrier when increasing the number of qubits. To prolong $T_2$ far beyond the timescale of quantum gate manipulations remains to be solved before further scalability can be addressed.

The coherence time, $T_2$, of NV$^-$ spin as measured by a Hahn echo sequence is determined by several contributions:
\begin{alignat}{2}
\frac{1}{T_{2}}{\;}\simeq{\;}&\left(\frac{1}{T_{2}}\right)_{\begin{subarray}{l} ^{13}\text{C}\\ \text{flip-flop} \end{subarray}}
+\left(\frac{1}{T_{2}}\right)_{\begin{subarray}{l} \text{nitrogen} \\ \text{impurity} \end{subarray}} 
+\left(\frac{1}{T_{2}}\right)_{\begin{subarray}{l} \text{paramag.} \\ \text{defect} \end{subarray}} \notag \\
&{\,}+\left(\frac{1}{T_{2}}\right)_{\begin{subarray}{l} \text{spin-lattice} \\ \text{relaxation} \end{subarray}}, \label{F1}
\end{alignat}
where the $T_2$ of NV$^-$ is dominated by spectral diffusion due to the fluctuation of local fields. These may be the result of nuclear spins ($^{13}$C) and/or nitrogen impurity spins (substitutional nitrogen in the neutral charge state, $\textrm{N}_\textrm{S}^0$, $S=1/2$) known as the P1 center in electron paramagnetic resonance (EPR) spectroscopy. The contribution of the last term is small even at RT since spin lattice relaxation is inefficient due to weak spin-orbit interaction of carbon and the strong bonding (i.e. high Debye temperature) of diamond. Carbon has two stable isotopes, $^{12}$C with $I=0$ (natural abundance of 98.9\%) and $^{13}$C with $I=1/2$ (1.1\%). In a natural abundance sample, $T_2$ of NV$^-$ at RT is $\sim0.65$~ms, limited by magnetic noise from the $^{13}$C nuclear spin bath.~\cite{Markham} By depleting the $^{13}$C content (99.7\% $^{12}$C enrichment) and increasing the purity ($[\textrm{N}_\textrm{S}^0]\sim0.05$~ppb), a $T_{2}$ of 1.8~ms has been attained for a \textit{native} NV$^-$ center (i.e. a {\it{grown-in}} NV$^-$ defect formed during crystal growth), in single crystal (SC) chemical vapor deposition (CVD) diamond.~\cite{Balasubramanian} It is thus desirable to use a $^{12}$C-enriched and high purity (low N concentration) diamond to achieve a long $T_2$. 

Nitrogen ion implantation into pure diamond is the primary technique used to fabricate NV$^-$ at a desired position.~\cite{Meijer,Gaebel,Rabeau,Naydenov1,Naydenov2,Pezzagna1,Schwartz,Pezzagna2,Pezzagna3,Toyli,Neumann2,Dolde} Yet the ion implantation process reduces the long coherence time drastically, presumably due to the presence of unpaired electrons from residual radiation damage. The $T_2$ of NV$^-$ produced by ion implantation and subsequent annealing is shorter (0.35~ms in Ref.~\onlinecite{Gaebel}) than the limit of natural abundance diamond (0.65~ms in Ref.~\onlinecite{Markham}), being dominated by the paramagnetic defect term in Eq.~(\ref{F1}).~\cite{Gaebel,Naydenov2} Even in $^{12}$C-enriched substrates, the $T_2$ of implanted NV$^-$ is at best $\sim0.5$~ms.~\cite{Dolde} Thus, the advantage of $^{12}$C enrichment has, to date, been unattainable for NV$^-$ fabricated by implantation. Importantly, a long $T_2$ of implanted NV$^-$ can be obtained only when the sources of spectral diffusion, deriving from residual defects, are significantly decreased in a high purity and $^{12}$C-enriched crystal. 

The irradiation of ions or electrons create vacancies and interstitials at RT. Self-interstitial defects (the EPR R1 and R2 centers) anneal out at 400-450$^{\circ}$C, while most ($\sim80$\%) of the created vacancies remain stable.~\cite{Twitchen,Iakoubovskii1} Vacancies, with a 2.3~eV activation energy of migration, become mobile at $\sim600^{\circ}$C.~\cite{Davies} Thus, an annealing temperature of $\sim800^{\circ}$C is usually employed for NV$^-$ fabrication by implantation. Thermal annealing has two roles, one to create NV through vacancy migration towards implanted nitrogen, and the other to remove unwanted residual defects. The short $T_2$ of implanted NV$^-$ suggests that the annealing at $\sim800^{\circ}$C leaves paramagnetic residual-defects that become the main source of decoherence.

Formation of paramagnetic defects by $^{17}$O ion implantation (100~MeV, $5\times10^{14}$~cm$^{-2}$, RT) and subsequent annealing were studied using {\it{ensemble}} EPR measurements in Ref.~\onlinecite{Iakoubovskii3}. Although the defect concentration dependence on the annealing temperature was not given in detail, various major paramagnetic defects were identified in the annealing range of 100-1400$^\circ$C.  Major paramagnetic defects in the annealing range of 850-1050$^\circ$C are $\langle110\rangle$ vacancy chains $\textrm{V}_\textit{n}^0$ ($n\geq3$, neutral charge, $C_{2v}$ symmetry) such as $\textrm{V}_3^0$ (R5), $\textrm{V}_4^0$ (O1), $\textrm{V}_5^0$ (R6), $\textrm{V}_6^0$ (R10), $\textrm{V}_7^0$ (R11) and $\textrm{V}_8^0$ (KUL11) where their concentrations decrease with increasing chain length. Two carbon dangling bonds, one at each end of the $n$-vacancy chains, give a spin $S=1$. The vacancy chains ($\textrm{V}_\textit{n}^0$, $n\geq3$), with a total concentration of $3.4\times10^{17}$~cm$^{-3}$ at $900^\circ$C, disappear after annealing at $T\geq1100^{\circ}$C except for a small fraction ($\sim1/10$) of R10 ($\textrm{V}_6^0$). At temperatures greater than $1100^\circ$C, paramagnetic multi-vacancy clusters of R10$'$ (remaining R10), R8, and R12 dominate, although $^{17}$O-related KUL12 ($S=1/2$) defects and preexisting KUL1 ([Si-V]$^0$, $S=1$) defects coexist.

Systematic studies on the temperature dependence of paramagnetic defect concentrations were reported in Ref.~\onlinecite{Lomer,Wilson}. Paramagnetic defects introduced by electron irradiation (2~MeV, $8\times10^{19}$~cm$^{-2}$, RT) and subsequent annealing (600-1400$^\circ$C) were measured by {\it{ensemble}} EPR in natural type-IIa diamond crystals. At $\sim600^{\circ}$C where vacancies are mobile, divacancy $\textrm{V}_2^0$ (R4/W6, $S=1$, $C_{2h}$ symmetry) anneals in. While $\textrm{V}_2^0$ anneals out at $\sim850^{\circ}$C, the formation of a series of $\langle110\rangle$ vacancy chains $\textrm{V}_\textit{n}^0$ such as $\textrm{V}_3^0$ (R5), $\textrm{V}_4^0$ (O1), $\textrm{V}_5^0$ (R6), $\textrm{V}_6^0$ (R10), $\textrm{V}_7^0$ (R11) starts to develop between $\sim750^{\circ}$C and $\sim950^{\circ}$C. 
Shorter chains are dominantly formed among the vacancy chains ($\textrm{V}_\textit{n}^0$, $n\geq3$), giving the chain length of $n=3.8$ on average. Each of dominant vacancy chains, $\textrm{V}_3^0$ (R5), $\textrm{V}_4^0$ (O1), and $\textrm{V}_5^0$ (R6), reach a maximum concentration ($2.3\times10^{18}$~cm$^{-3}$ in total) at $1000^\circ$C and decrease with increasing temperature ($T>1000^{\circ}$C). Instead, multi-vacancy clusters having a different configuration to the chain-type, labeled as R7, R8 and R12,~\cite{Lomer,Iakoubovskii2,Baker} start to appear at $T>1000^{\circ}$C. The R12 center ($C_{3v}$ symmetry), being stable even at $\sim1400^{\circ}$C, is attributed to a multi-vacancy-cluster of ring type.~\cite{Lomer}  It was shown in Ref.~\onlinecite{Lomer} that the concentration of residual paramagnetic defects is minimized at $\sim1250^{\circ}$C.

Naydenov $\textit{et al.}$ reported the effect of high-temperature annealing on $T_2$ of implanted NV$^-$, where single NV$^-$ centers were fabricated by N ion implantation ($^{15}$N$^+$, 150 and 300~keV, $1\times10^8$~cm$^{-2}$) and annealed at two different temperatures, 800$^\circ$C and 1200$^\circ$C.~\cite{Naydenov2} The comparison study showed annealing at 1200$^\circ$C has a positive effect, which was observed as an increase in the proportion of NV centers with a $T_2\geq50$~$\mu$s. However, the $T_2$ of implanted NV$^-$ remained shorter than the best (0.65~ms) of {\it{native}} NV$^-$ in natural abundance.

Here we report a long $T_2$ up to 2~ms of single NV$^-$ centers by implantation after annealing at $1000^\circ$C, using a $^{12}$C enriched and high-purity diamond. To identify the source of decoherence, the thermal behaviors of residual defects created by nitrogen implantation were also studied by {\it{ensemble}} EPR. We find that a substantial decrease of the concentration of paramagnetic residual defects occurs at a temperature of $\sim1100^{\circ}$C, rather lower than that of $\sim1250^{\circ}$C (Ref.~\onlinecite{Lomer} for electron irradiation). This paper thus consists of two parts: the first aims at achieving long $T_2$ times of individual implanted-NV$^-$ centers and the second is to study the residual defects by {\it{ensemble}} measurements, where two different fluences of nitrogen implantation were used according to each purpose. The samples used for {\it{ensemble}} EPR measurements have been also characterized by photoluminescence (PL) spectroscopy. Lastly, the long $T_2$ observed here is discussed on the basis of the {\it{ensemble}} EPR measurements for residual defects. 

\section{Experiments}
\subsection{Single NV$^-$ centers fabricated by low fluence and {\itshape{ex-situ}} annealing}
High-purity 99.99\% $^{12}$C enriched SC CVD diamond was used in order to achieve a long $T_2$. The sample was implanted with a 10~MeV $^{15}$N$^{3+}$ micro-beam with a full-width at half-maximum (FWHM) of 1.5~$\mu$m (the micro-beam system connected with the 3~MV Tandem accelerator at JAEA, Takasaki~\cite{Sakai,Kamiya}) at RT in a vacuum of $\sim10^{-6}$~Torr. The incident beam was scanned with $\sim3$ ions per site ($2\times10^8$~ions/cm$^2$ on average) to form a square grid of implantation sites separated by $\sim8$~$\mu$m in the area of $200\times200$~$\mu$m$^2$ ($5\times10^6$~ions/cm$^2$ on average). This low fluence allows creation of single NV centers which can be characterized individually by confocal microscopy.~\cite{Gruber} To create NV centers, the sample was annealed at $1000^{\circ}$C for 2~h in a vacuum of 3$\times$10$^{-6}$~Torr. The resultant NV centers were observed using a home-built confocal microscope with a 532~nm excitation laser. Continuous-wave and pulsed optically-detected magnetic resonance (ODMR) were measured for single NV$^-$ centers under a static magnetic field at RT.

\subsection{{\itshape{Ensemble}} measurements using high fluence and {\itshape{in-situ}} annealing}
For the low fluence $(\sim1\times10^{6}$~/cm$^2)$ used for fabrication of single NV$^-$ centers, the concentration of paramagnetic residual defects is far below the detection limit of {\it{ensemble}} measurements of conventional EPR. Also, individual weakly luminescent defects like vacancy clusters are unable to be measured by confocal microscopy. To obtain information for residual defects, the another sample with different implantation conditions was used for the {\it{ensemble}} EPR measurements: high-purity natural abundance CVD plates (Element Six Ltd., electronic grade, [Ns$^0$]$\sim$0.6-0.8~ppb in bulk measured by EPR, $4.5\times4.5\times0.5$~mm$^3$, $(100)$-oriented single crystal).  The samples were implanted with N ions at seven different energies between 4 and 13~MeV in a vacuum of $\sim10^{-7}$~Torr at elevated temperatures from $800^{\circ}$C to $1200^{\circ}$C. The total fluence was $1.4\times10^{14}$~cm$^{-2}$ since N ions of $1\times10^{13}$~cm$^{-2}$ for each energy were implanted at both sides of sample surface. The use of high fluence allows lower sensitivity {\it{ensemble}} EPR measurements to be performed. In fact the fluence was chosen to be below $\sim1$~ppm for the average N concentration in the implanted layer. The number of vacancies created by the implantation is expected to be $\sim10^{20}$~vacancies/cm$^3$ by SRIM Monte Carlo simulation,~\cite{Ziegler} far below the graphitization threshold ($\sim10^{22}$~vacancies/cm$^3$).~\cite{Uzan} In the case of 1.7~MeV N$^+$ implantation ($10^{16}$~cm$^{-2}$) into type IIa diamond crystal, the spin density of the EPR signal arising from amorphous carbon after {\it{in-situ}} annealing at 1000$^{\circ}$C was lower by an order of magnitude than {\it{ex-situ}} annealing.~\cite{Lee} In the fluence used in our present work, damage cascades could overlap ($1\times10^{13}$~cm$^{-2}\sim1/(3~\text{nm})^{2}$). So, we employed {\it{in-situ}} annealing in order to prevent excess vacancies from accumulating and thus a high degree of crystallinity was preserved. The samples were firstly heated to $700^{\circ}$C over a few minutes, and then set to the target temperature with a heating rate of $\sim10^{\circ}$C/min. The {\it{in-situ}} annealing temperatures were kept for $\sim3$~h while changing the beam-transport with different energies. Promptly after implantations, samples were furnace-cooled to RT for 1~h. The thermal behaviors of residual defects were characterized by {\it{ensemble}} measurements of EPR. Continuous-wave EPR measurements were performed on a Bruker BioSpin E500 X-band spectrometer. Identification of paramagnetic defects by {\it{ensemble}} EPR was based on angular dependence of the line positions. To align the crystal orientation with the EPR signals arising from symmetry-related sites at RT, more than $\sim$2$\times$10$^{12}$ spins were required for a typical defect in diamond. The samples  have been also characterized by PL (photoluminescence). The PL spectra were measured at 83~K with an excitation power of a few tenths of a $\mu$W using a home-built spectrometer. 

\section{Results}
\subsection{Spin coherence times of single NV$^-$ centers}

\begin{figure}
\includegraphics[width=8.6cm]{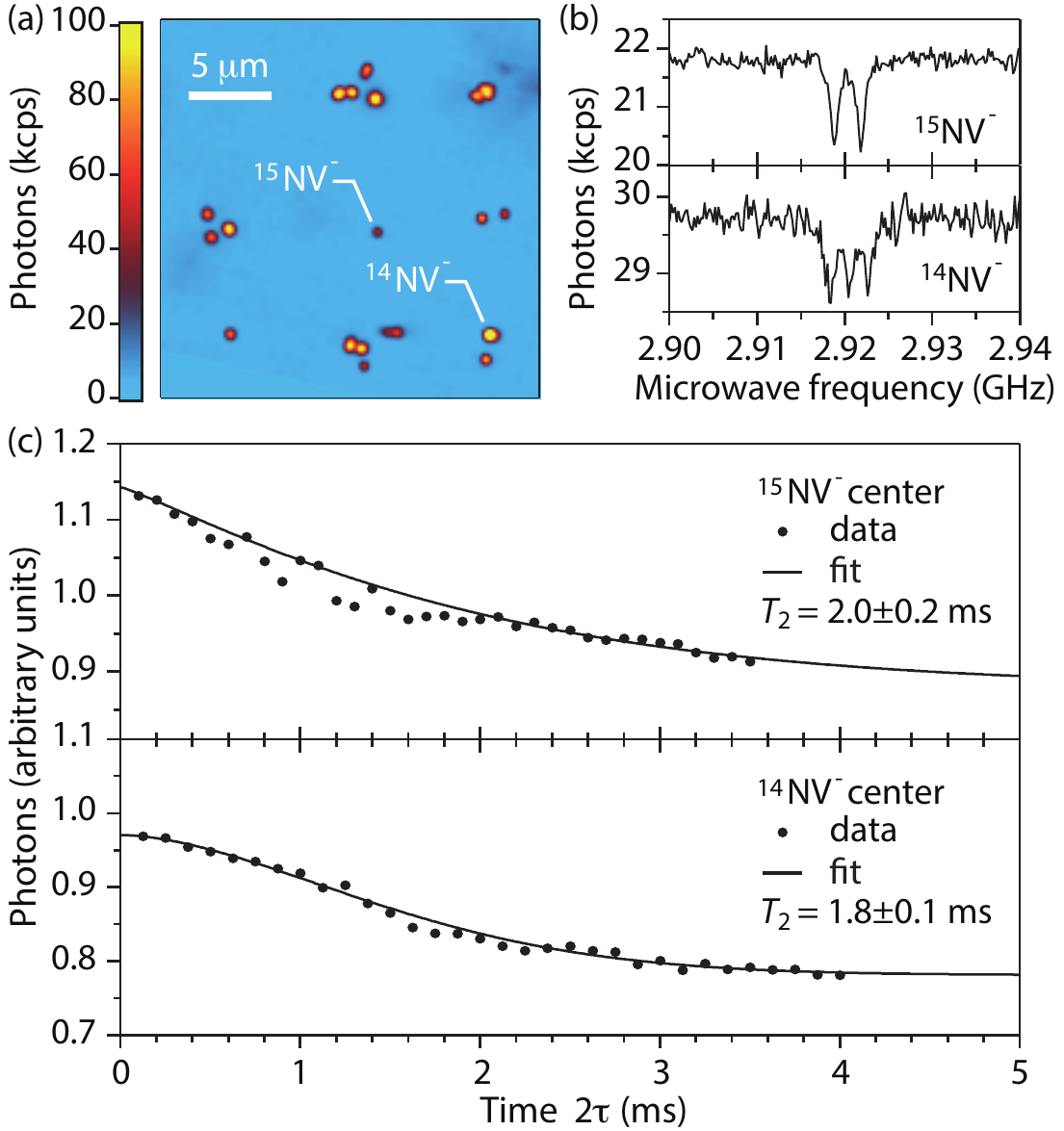}%
\caption{(Color online) Room temperature characteristics of single NV$^-$ centers fabricated in high purity, 99.99\% enriched SC CVD diamond by micro-beam $^{15}$N$^{3+}$ (10~MeV) implantation at RT and subsequent annealing at 1000$^\circ$C in vacuum. (a) Confocal microscope image of NV centers (yellow/red dots) with 532~nm excitation. (b) ODMR spectra of $^{15}$NV$^-$ with the hyperfine splitting $A=3.1$~MHz, and $^{14}$NV$^-$ with $A = 2.2$~MHz. (c) The echo decay curves of $^{15}$NV$^-$ (top) and $^{14}$NV$^-$ (bottom) measured by a Hahn echo sequence. The solid lines indicate the fit function $E(2\tau){\propto}{\exp}[-(2\tau/T_{2})^{\alpha}]$, where the index $\alpha=1.1$ for $^{15}$NV$^-$ and $\alpha=1.8$ for $^{14}$NV$^-$.  \label{FIG1}}
\end{figure}

Single NV$^-$ centers in 99.99\% $^{12}$C enriched SC CVD diamond by micro-beam implantation at RT followed by annealing at 1000$^\circ$C were characterized by scanning confocal microscopy and optically detected magnetic resonance (ODMR).  Nitrogen has two stable isotopes, $^{14}$N ($I=1$, natural abundance 99.63\%) and $^{15}$N ($I=1/2$, 0.37\%). The ODMR of either the $^{15}$N hyperfine structure or $^{14}$N allowed determination of whether the observed NV$^-$ centers were created by implanted ($^{15}$N) or pre-existing impurities ($^{14}$N) in the substrate.~\cite{Rabeau} The ODMR spectra of single $^{14}$NV$^-$ and $^{15}$NV$^-$ centers in the confocal image (Fig.~\ref{FIG1}(a)) are shown in Fig.~\ref{FIG1}(b). The triplet and doublet hyperfine structures observed in the ODMR spectra between $|{\it{m}}_{\rm{s}}=0\rangle$ and $|{\it{m}}_{\rm{s}}=+1\rangle$ indicate a single $^{14}$NV$^-$ with the hyperfine constant $A=2.2$~MHz (bottom) and a single $^{15}$NV$^-$ with $A=3.1$~MHz (top), respectively (Fig.~\ref{FIG1}(b)). The both $^{14}$NV$^-$ and $^{15}$NV$^-$ centers were located at a depth of $\sim3.8$~$\mu$m, which agrees well with the projected ion-stopped range of 3.82~$\mu$m by SRIM Monte Carlo simulation.~\cite{Ziegler} We also found that the $^{14}$NV$^-$ and $^{15}$NV$^-$ centers were observed at the square grid points with separations of $\sim8$~$\mu$m within in-plane radius $<1.4~\mu$m. Hence, we conclude that the both $^{14}$NV$^-$ and $^{15}$NV$^-$ are not native but created by $^{15}$N ion implantation.  The $^{14}$NV$^-$ was created through the trapping of a vacancy by a pre-exiting $^{14}$N atom during annealing, while the $^{15}$NV$^-$ was from the implant. Detailed analysis of the yield of NV$^-$ from the two different nitrogen sources will be given elsewhere.~\cite{Yamamoto}

Figure~\ref{FIG1}(c) shows the echo decay curves of the single $^{15}$NV$^-$ center (top) and $^{14}$NV$^-$ center (bottom) taken by pulsed ODMR spectroscopy using a Hahn echo sequence.~\cite{Gaebel}  By fitting the data with $E(2\tau){\propto}{\exp}[-(2\tau/T_{2})^{\alpha}]$ where $\alpha$ used as the free parameter, the spin coherence time $T_2$ was measured to be $2.0{\pm}0.2$~ms for the $^{15}$NV$^-$ and $1.8{\pm}0.1$~ms for the $^{14}$NV$^-$. Thus, two NV centers consisting of different nitrogen sources have similar $T_2$ of $\sim$2~ms.
To confirm the reproducibility of long $T_2$, we have measured another 99.99\% $^{12}$C-enriched crystal with similar implantation and annealing, and then obtained long $T_2$ of up to 1.6~ms in implanted NV centers (data not shown). 

\subsection{{\itshape{Ensemble}} measurements of thermal behaviors} 
\begin{figure}
\includegraphics[width=8.6cm]{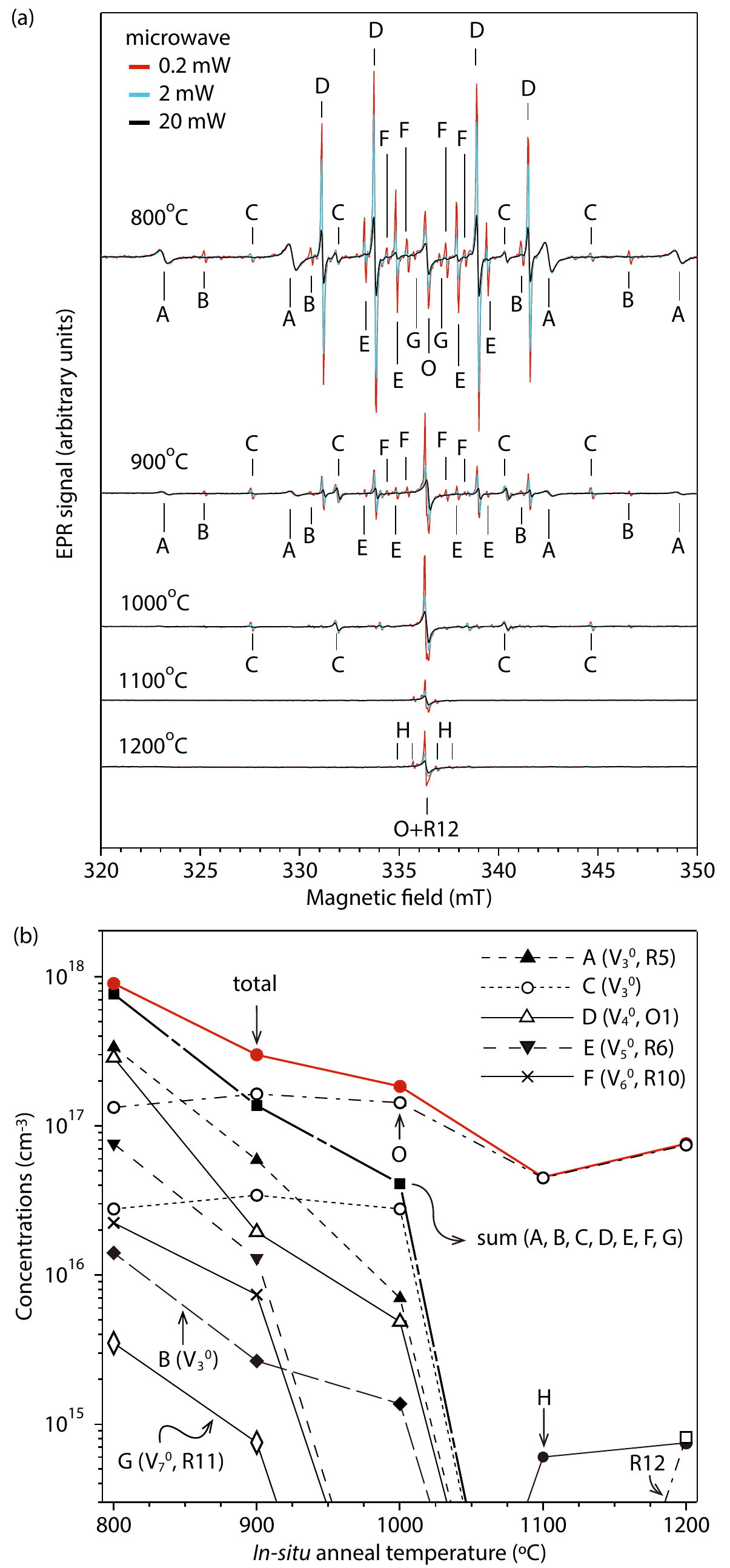}%
\caption{(Color online) (a) {\it{Ensemble}} EPR spectra (${\mathbf{B}}$$\parallel$$[100]$, RT, in dark, 9.428~GHz) of high-purity natural abundance SC CVD diamond plates implanted with N ions (1$\times$10$^{13}$~cm$^{-2}$ at each of 7 steps between 4 and 13~MeV) at various \textit{in-situ} annealing temperatures (800 to 1200$^{\circ}$C). Multi-vacancy chain centers ({\textquoteleft}A' to {\textquoteleft}G') and other defects ({\textquoteleft}O', {\textquoteleft}H', and R12) were observed. (b) The dependence of the concentration of paramagnetic centers on \textit{in-situ} annealing temperature. {\textquoteleft}Total' was estimated by integrating the whole EPR spectra which includes all signals. The concentrations were calculated by assuming that the defects are formed inside 4.85~$\mu$m-thick layers in both sides of the samples, where 4.85~$\mu$m corresponds to the projected ion stop range for 13~MeV N ion implantation.\label{FIG2}}
\end{figure}

The long $T_2$ of single NV$^-$ as shown above indicates that a substantial decrease of paramagnetic residual defects is achieved by employing annealing temperature of 1000$^\circ$C. We have thus examined the thermal behaviors of residual defects in the case of N ion implantation in details. Figure~\ref{FIG2}(a) shows the EPR spectra (${\mathbf{B}}$$\parallel$$[100]$, taken at RT) of N implanted high-purity SC CVD diamond plates with various {\it{in-situ}} annealing temperatures (800 to 1200$^{\circ}$C). Ten EPR signals ({\textquoteleft}A' to {\textquoteleft}G', {\textquoteleft}O', {\textquoteleft}H', and {\textquoteleft}R12') were distinguished as indicated in Fig~\ref{FIG2}(a). The signals of {\textquoteleft}A' to {\textquoteleft}G' exhibited a common quartet-line signature (two distinguishable sites with the intensity ratio of 1:2, each consisting of two fine-structure lines), which coincides with the features of the $\langle110\rangle$ multi-vacancy chain with a spin $S=1$. We have checked their angular patterns in magnetic-field rotation experiments from ${\mathbf{B}}$$\parallel$$[100]$ to ${\mathbf{B}}$$\parallel$$[011]$. By comparing our angular-pattern data with the literature,~\cite{Iakoubovskii3} the defects {\textquoteleft}A', {\textquoteleft}D', {\textquoteleft}E', {\textquoteleft}F' and {\textquoteleft}G' were identified as V$_3$$^0$ (R5), V$_4$$^0$ (O1), V$_5$$^0$ (R6), V$_6$$^0$ (R10), V$_7$$^0$ (R11), respectively. In addition, the observed fine structure splitting in V$_n$$^0$ with $n\geq4$ agreed well with that calculated using point-dipole approximation.~\cite{Lomer,Iakoubovskii3} We also found that the fine structure splitting of the defect {\textquoteleft}B' was almost same as the R5 center at {\it{low temperature}} ($\leq$77~K).~\cite{Iakoubovskii3} Thus, the defect {\textquoteleft}B' is probably a similar defect of R5 ({\textquoteleft}A') where the dynamics causing the temperature dependence in the R5 ({\textquoteleft}A') might be frozen at RT in the defect {\textquoteleft}B'. In Ref.~\onlinecite{Lomer}, it is reported that a deviation of the fine structure splitting of V$_3$$^0$ (R5) from a theoretical one using a point-dipole approximation is attributed to a delocalization of the wave function, which is not negligible for the short chain length. It is also reported in the Ref.~\onlinecite{Lomer} that there are a few different centers ascribed to V$_3$$^0$. Therefore, we speculate the defects of {\textquoteleft}B' and {\textquoteleft}C' to be analogs of V$_3$$^0$. All signals assigned to multi-vacancy chains ({\textquoteleft}A' to {\textquoteleft}G') except for {\textquoteleft}C' decreased with increasing {\it{in-situ}} annealing temperature from 800$^{\circ}$C to 1000$^{\circ}$C as can be seen in Fig.~\ref{FIG2}(a). Note that the EPR spectrum from V$_2$$^0$ (R4/W6)~\cite{Twitchen2} was not observed at both RT and 32~K in this study. At temperatures $T\geq1100^{\circ}$C, multi-vacancy chains ({\textquoteleft}A' to {\textquoteleft}G') were annealed out and different types of defects were observed, as denoted by {\textquoteleft}H' and {\textquoteleft}R12' in Fig.~\ref{FIG2}(a). The R12 center ($C_{3v}$ symmetry, $S = 1$, overlapping to {\textquoteleft}O') is reported to be a typical defect appearing in the annealing stages above 1100$^{\circ}$C.~\cite{Iakoubovskii3} On the other hand, we tentatively assign that the signal {\textquoteleft}H' is a multi-vacancy cluster without having the $C_{2v}$ symmetry of vacancy chain structures. The central broad signal {\textquoteleft}O' is isotropic and may arise from more than two species, probably including surface damage such as amorphous carbon as well. At 1200$^{\circ}$C, the signal {\textquoteleft}O' turns from decreasing to increasing, suggesting that the annealing products from multi-vacancy chains may contribute to the defect formation. In fact, the previous work suggested that multi-vacancy chains are converted into a ring-type configuration such as the hexagonal vacancy cluster, being stable up to 1400$^{\circ}$C.~\cite{Iakoubovskii2,Iakoubovskii3}

\begin{figure}
\includegraphics[width=8.6cm]{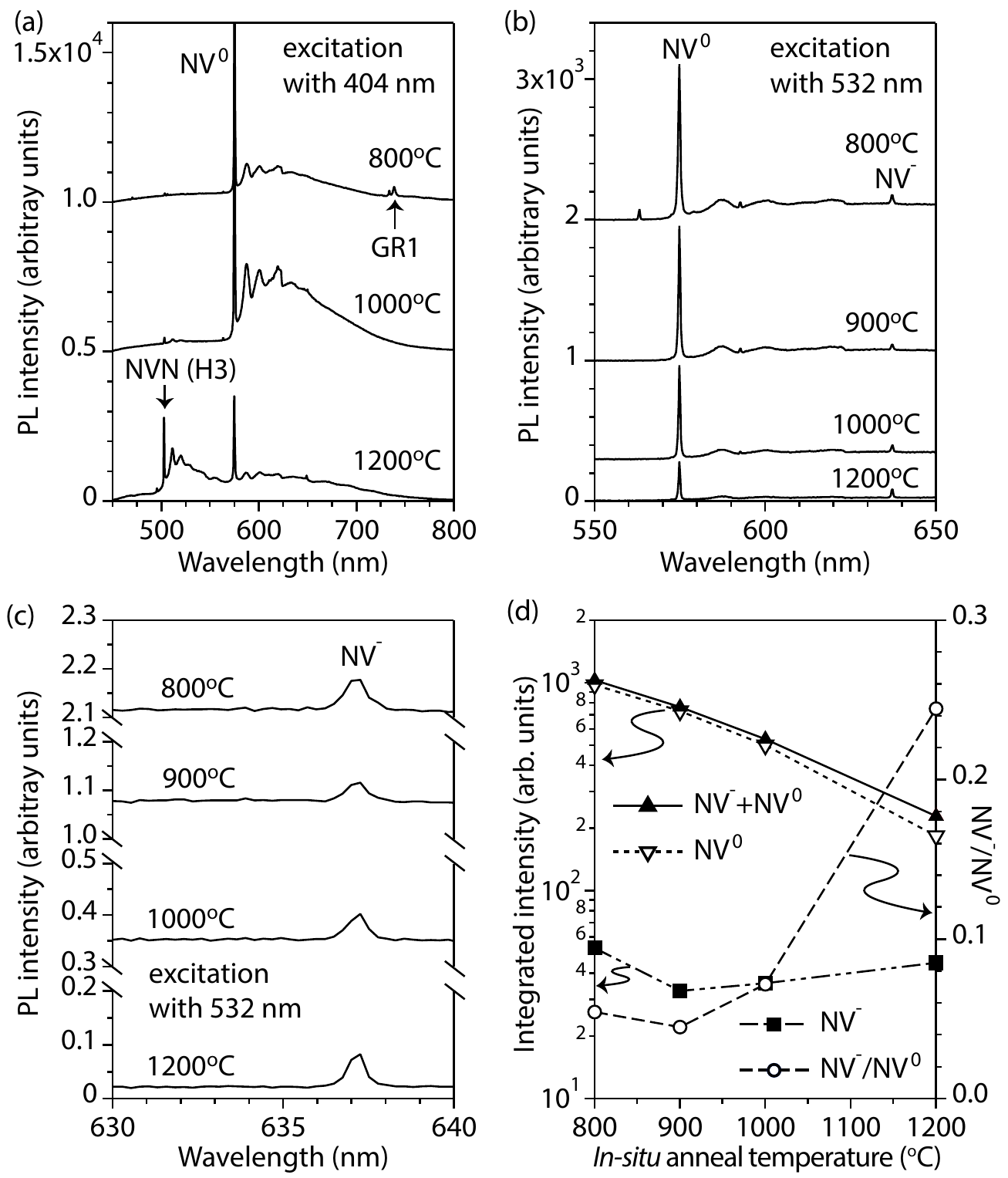}%
\caption{PL spectra (83~K) of high-purity natural abundance SC CVD diamond plates implanted with N ions (1$\times$10$^{13}$~cm$^{-2}$ at each of 7 steps between 4 and 13~MeV) at various {\it{in-situ}} annealing temperatures (800 to 1200$^\circ$C) with (a) 404~nm and (b) 532~nm excitation laser. (c) The enlarged view of (b) in the range between 630 and 640~nm. (d) The integrated intensity of NV$^0$ (inverted triangles), NV$^-$ (squares), ${\text{NV}}^{0}+{\text{NV}}^{-}$ (triangles), and the ${\text{NV}}^{-}/{\text{NV}}^{0}$ ratio (circles).\label{FIG3}}
\end{figure}

The dependence of concentrations of the defects on the \textit{in-situ} annealing temperature is shown in Fig.~\ref{FIG2}(b). The most dominant defect was associated with the broad signal {\textquoteleft}O'. This signal was reduced to 5\% at $1100^{\circ}$C as compared to that at $800^{\circ}$C. The total concentration of the multi-vacancy clusters ({\textquoteleft}A', {\textquoteleft}B', {\textquoteleft}C', {\textquoteleft}D', {\textquoteleft}E', {\textquoteleft}F', {\textquoteleft}G', {\textquoteleft}H', R12) was minimized at $\sim1050^{\circ}$C and at least three orders of magnitude less than that at $800^{\circ}$C. In the case of ion implantation, in both {\it{in-situ}} (present work) and {\it{ex-situ}} (Ref.~\onlinecite{Iakoubovskii3}) annealing, multi-vacancy chains disappear at $T\sim1050^{\circ}$C and multi-vacancy cluster of ring-type appear at $\sim1100^{\circ}$C. Thus, the total concentration of the paramagnetic residual defects (red line) was minimized at 1100$^{\circ}$C, rather lower than that of $\sim1250^{\circ}$C for electron irradiation (Ref.~\onlinecite{Lomer}). The number of the paramagnetic defects can be estimated as 3~spins/ion at $800^\circ$C, 1~spin/ion at $900^\circ$C, and less than 1~spin/ion at $T\geq1000^\circ$C. 

The samples used for {\it{ensemble}} EPR measurements have been also characterized by photoluminescence (PL) spectroscopy. Figure~\ref{FIG3}(a) shows 404~nm excited PL spectra observed at 83~K. The GR1 peak that originates from neutral vacancy (V$^0$) and has a zero phonon line (ZPL) at 741~nm was observed in the sample implanted at $800^{\circ}$C. The GR1 peak was not present for implantation temperatures of above $1000^{\circ}$C, while the H3 peak (ZPL at 503~nm) originating from the (N-V-N)$^0$ center appeared in the sample implanted at $1200^{\circ}$C. Note that the temperature of $1200^{\circ}$C is much lower than that ($1700^{\circ}$C) for formation of the A center (N-N) by N diffusion during a high-pressure high-temperature (HPHT) treatment in type-Ib crystals.~\cite{Chrenko} It is also reported that the A center may be formed by heating at $1500^{\circ}$C {\it{in vacuo}} after electron irradiation.~\cite{Collins}
In order to obtain information on NV centers, PL measurements at 83~K using 532~nm excitation were carried out {\big(}Fig.~\ref{FIG3}(b) and ~\ref{FIG3}(c){\big)}. Both NV$^0$ (ZPL at 575~nm) and NV$^-$ (ZPL at 637~nm) peaks were observed in all samples. The integrated intensity of NV$^0$ and NV$^-$ lines as well as their ratios extracted from Fig.~\ref{FIG3}(b) are plotted in Fig.~\ref{FIG3}(d). We found that while the total intensity of $\rm{NV}^{-}+\rm{NV}^0$ decreased with increasing implantation temperature, the NV$^-$/NV$^0$ ratio increased with a rise in temperature.

\section{Discussion}
In our present work, coherence times ($T_2$) of single NV$^-$ centers fabricated by 10~MeV $^{15}$N ion micro-beam implantation have been measured. A long $T_2$ of 2~ms was obtained by selecting an {\it{ex-situ}} annealing temperature of 1000$^{\circ}$C and the use of a high-purity, isotopically pure ($^{13}$C-0.01\%, $^{12}$C-99.99\%) substrate. For implanted NV$^-$ in $^{13}$C-0.01\% SC CVD diamond, $T_2$ times of 0.1~ms (18~MeV N implantation, Ref.~\onlinecite{Neumann2}) and 0.5~ms (1~MeV, Ref.~\onlinecite{Dolde}) have previously been reported for NV$^-$ pairs. In these studies, the samples were annealed at 800$^\circ$C for 2~h in Ref.~\onlinecite{Neumann2} and 8~h in Ref.~\onlinecite{Dolde}, respectively. Comparison with these $T_2$ times suggests that a modest increase of annealing temperature from 800$^\circ$C to 1000$^\circ$C prolongs $T_2$ significantly. The $T_2$ time in the present work far exceeds the best (0.65~ms) of {\it{native}}-NV in natural abundance ($^{13}$C-1.1\%) diamond, and reaches the longest reported so far in $^{13}$C-depleted samples.~\cite{Balasubramanian,Ishikawa,Jahnke}
The annealing dependence of residual paramagnetic defects has been studied by using {\it{in-situ}} annealed natural abundance SC CVD diamonds implanted with N ions of a high fluence ($1\times10^{13}$~cm$^{-2}$ at each of 7 steps between 4 and 13~MeV). These implantation conditions are well suited to {\it{ensemble}} EPR measurements. The {\it{ensemble}} EPR measurements have elucidated that residual defects remain at significant concentrations with implantation temperatures up to 800$^\circ$C while they substantially decrease in the 800$^\circ$C to 1100$^\circ$C range. Above 1100$^\circ$C, paramagnetic defects having configurations stable at higher temperatures appear. The {\it{ensemble}} EPR measurements suggest that improvement in $T_2$ of single NV$^-$ centers by employing an {\it{ex-situ}} annealing temperature of 1000$^\circ$C has been attained due to substantial reduction of paramagnetic defects and therefore the third term in Eq.~\ref{F1}.
\begin{figure}
\includegraphics[width=8.6cm]{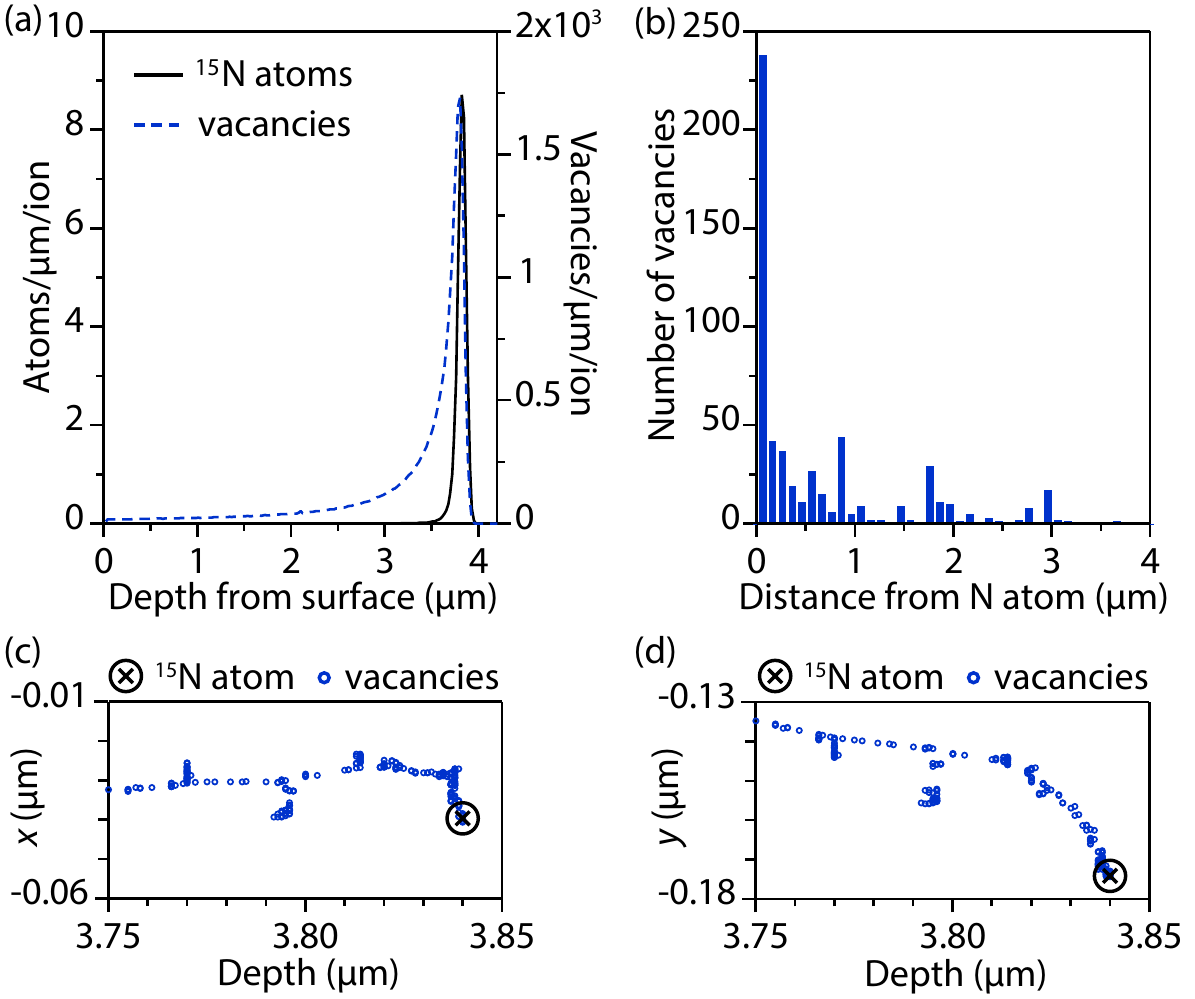}%
\caption{(Color online) SRIM Monte Carlo simulation of spatial distribution for $^{15}$N atoms and vacancies created by 10~MeV $^{15}$N ion implantation. (a) Statistical depth distribution of implanted $^{15}$N atoms (black solid line) and vacancies (blue dashed line) obtained as the average of a total number of $8\times10^4$ incident ions. (b) Distance from individual vacancies to $^{15}$N atom for a typical collision cascade created by a single 10~MeV $^{15}$N ion. Vacancies (blue circle) near an implanted $^{15}$N atom (black cross) for a collision cascade are shown in (c) the {\it{x-z}} projection plane and in (d) the {\it{y-z}} plane, where {\it{z}} implantation direction. Here, 42\% of the total number of 559~vacancies are plotted on both figures of (c) and (d). \label{FIG4}}
\end{figure}

Now, we consider the formation of paramagnetic vacancy clusters in the vicinity of implanted NV$^-$. The statistical depth distributions of implanted $^{15}$N atoms (black solid line) and vacancies (blue dashed line) as computed by SRIM code (Stopping and Range of Ions in Matter,~\cite{Ziegler} with diamond density of 3.52~g/cm$^3$, displacement energy of 37.5~eV,~\cite{Koike} and a total number of $8\times10^4$ incident 10~MeV $^{15}$N ions) are shown in Fig.~\ref{FIG4}(a). The average depth of individual $^{15}$N atoms is calculated to be $\sim3.82$~$\mu$m. The vacancy distribution due to $^{15}$N implantation peaks at $\sim3.80$~$\mu$m, and 43\% of vacancies among the average number of $\sim570$~vacancies/ion are produced within a depth range of $\pm100$~nm from the center of the peak. On the other hand, the distance of individual vacancies from a $^{15}$N atom for a typical collision cascade is shown in Fig.~\ref{FIG4}(b). Again, 40\% of vacancies are created in a small volume, within a distance of 100~nm from the implanted $^{15}$N atom. For high energy implantation, such a large number of vacancies near the implanted N atom contribute to the high NV$^-$ formation yield.~\cite{Meijer,Pezzagna1} Interactions with the carbon atoms in the target sample cause both carbon displacements from lattice sites and the trajectory change of the implanted $^{15}$N atom. Secondary knock-ons can also result in several branches of collision cascade in the vicinity of the ion track as shown in Fig.~\ref{FIG4}(c) ({\it{x-z}} projection plane) and Fig.~\ref{FIG4}(d) ({\it{y-z}} plane) for a single 10~MeV $^{15}$N ion. Thus, vacancies are created along the $^{15}$N ion track and the branches of recoiling carbon atoms, which is favorable to the formation of vacancy clusters after annealing. The resultant vacancy clusters near the NV$^-$ are the main source of decoherence of NV$^-$ as observed in Ref.~\onlinecite{Dolde,Neumann2} (800$^{\circ}$C anneal). We attribute the long $T_2$ found in the present work as being essentially due to a reduction of paramagnetic vacancy clusters near NV$^-$ by annealing at 1000$^{\circ}$C.

A number of processes that reduce the concentration of paramagnetic vacancy clusters take place on annealing. The divacancy ($\textrm{V}_2^0$) and multi-vacancy chains ($\textrm{V}_n^0$, $n\geq3$) are paramagnetic, whereas the mono-vacancy of the neutral charge state ($\textrm{V}^0$) is diamagnetic. One possible process to reduce the concentration of $\textrm{V}_n^0$ ($n\geq2$) is the aggregation of vacancies by converting smaller clusters into larger ones. In this process, the total concentration of unpaired electron spins decreases even if the total number of vacancies involved is constant. In our {\it{ensemble}} EPR studies for samples formed by {\it{in-situ}} annealing during high fluence implantation, the divacancy was already annealed out by $800^\circ$C. All of the multi-vacancy chains ($\textrm{V}_n^0$, $n=3$ to 7) decreased with increasing the {\it{in-situ}} annealing temperature from $800^\circ$C to $1000^\circ$C, except the concentration of {\textquoteleft}C' assigned tentatively to an analog of $\textrm{V}_3^0$ stayed nearly constant. Thus, aggregation by increasing the chain length of linear vacancy chains is not the dominant process in this temperature range. It was pointed out that the total number of vacancies involved in the multi-vacancy chains was $\sim1/10$ of vacancies produced initially, which was estimated from GR1 ($\textrm{V}^0$) optical absorption measurements.~\cite{Lomer,Iakoubovskii3} This suggests that while a small part of mono-vacancies contributed to formation of the paramagnetic multi-vacancy chains formed by {\it{ex-situ}} annealing, a large fraction were annihilated by the recombination of vacancies with carbon interstitials or forming the diamagnetic vacancy clusters (see also Ref.~\onlinecite{Laszlo}).~\cite{Iakoubovskii3} Similarly, these processes might cause the substantial reduction of concentrations of the multi-vacancy chains by {\it{in-situ}} annealing ($T>800^\circ$C) in the present study. We speculate that such thermal behavior should also dominate in the case of {\it{ex-situ}} anneal after low fluence implantation thereby allowing a significant reduction of the paramagnetic defect concentration and a significant increase in $T_2$ particularly with a 1000$^\circ$C anneal.

Next, we consider the difference of the two fluences used in the present study from the viewpoint of the role of mobile vacancies in formation of paramagnetic vacancy clusters. There is controversy in the vacancy diffusion lengths determined experimentally.~\cite{Orwa,Santori} Here, we use a theoretical estimation.~\cite{Orwa} The low fluence was required to fabricate singe NV$^-$ centers. In this experiment, the vacancy diffusion length $L=\sqrt{6Dt}$ is estimated to be 130~nm, where the diffusion coefficient is $D=D_0\exp[-E_{\text{a}}/(k_{\text{B}}T)]$, with $D_0=3.7\times10^{-6}$~cm$^{2}$/s (Ref.~\onlinecite{Hu}), $E_{\text{a}}=2.3$~eV (Ref.~\onlinecite{Davies}), and $T=1000^{\circ}$C, and the annealing time is $t=7200$~s. $L$ is comparable to the averaged in-plane straggling length of 10~MeV $^{15}$N ions ($2\sigma\sim180$~nm, where $\sigma$ is the standard deviation of the Gaussian fit) and considerably shorter than the spot size of the micro-beam (FWHM of 1.5~$\mu$m).  When $\sim3$ ions are implanted at each implantation site as employed in this study, the distance between $^{15}$N atoms should be larger than $L$ for most of cases. Note that the probability of nitrogen atoms located within $\sim$300~nm can be estimated to be $\sim$1\%. Thus, it is likely that NV$^-$ and vacancy clusters are formed from vacancies resulting from the damage cascade caused by the implanted $^{15}$N ion. On the other hand, a high fluence ($1\times10^{13}$~cm$^{-2}$ at each of 7 steps of implantation energy) was required to perform the {\it{ensemble}} measurement of EPR. Since the high fluence can cause overlapping of the vacancy profiles of neighboring implants, {\it{in-situ}} annealing was employed to prevent vacancies from accumulating excessively. The diffusion length $L$ is calculated to be $\sim160$~nm ($T=1000^{\circ}$C, $t=10800$~s), which is longer than the separations of implanted N atoms (10-70~nm). As a result the vacancies resulting from individual ion impacts may overlap with neighboring implant spots to form vacancy clusters. In addition to dynamic annealing processes this may result in a difference in the formation of vacancy clusters between the low- and high-fluence experiments used in the present study. As a result, the annealing temperature which minimizes the paramagnetic residual defects might be slightly different for {\it{ex-situ}} annealing after low fluence implantation as used here for single NV$^-$ center formation, than {\it{in-situ}} annealing during high fluence implantation employed to fabricate samples for {\it{ensemble}} EPR studies. For high fluence, we note that similar thermal behavior has been observed between {\it{in-situ}} (this work) and {\it{ex-situ}} (Ref.~\onlinecite{Iakoubovskii3}) annealing: multi-vacancy chains disappeared at $T\sim1050^{\circ}$C and multi-vacancy cluster of ring-type appeared at $T\sim1100^{\circ}$C. 

The isotopic purity ($^{13}$C-0.01\%) of our diamond is expected to contribute greatly to the long coherence times observed. However, further improvement may still be possible when compared the $T_2$ time of 1.8~ms for a {\it{native}} NV$^-$ in $^{13}$C-0.3\%.~\cite{Balasubramanian} In Ref.~\onlinecite{Balasubramanian}, the concentration of isolated substitutional nitrogen ($\textrm{N}_\textrm{S}^0$) was estimated to be 0.05~ppb ($\simeq9\times10^{12}$~cm$^{-3}$) by extrapolating the correlation of the concentration of $\textrm{N}_\textrm{S}^0$ and that of NV$^-$ observed in various CVD samples (Ref.~\onlinecite{Tallaire}). Jahnke $\textit{et al.}$ reported that a {\it{native}} NV exhibited a $T_2$ of 2~ms in $^{13}$C-0.002\% polycrystalline CVD diamond.~\cite{Jahnke}  This $T_2$ is expected to be dominated by the paramagnetic nitrogen ($S=1/2$) term in Eq.~(\ref{F1}), since the N concentration was determined to be $[\textrm{N}_\textrm{S}^0]\sim4$~ppb ($\simeq7\times10^{14}$~cm$^{-3}$) by $\it{ensemble}$ EPR measurement. Given the similar $T_2$ between Ref.~\onlinecite{Jahnke} and ours, the concentration of paramagnetic defects in our sample is estimated to be a similar value of 4~ppb ($\simeq1/(110~\text{nm})^3$). Thus, the 2~ms $T_2$ time of implanted NV$^-$ we obtain in 99.99\% $^{12}$C enriched diamond is still determined by the paramagnetic defect term in Eq.~\ref{F1}. From the {\it{ensemble}} EPR measurements of {\it{in-situ}} annealing of high fluence implantation, the total concentration of the paramagnetic residual defects was found to be minimized at $\sim1100^{\circ}$C. The route explored in this work to fabricate qubits with long coherence times may show some improvement with further optimization with annealing temperatures between $1000^{\circ}$C and $1100^{\circ}$C.  

\section{Conclusion}
The achievement of long $T_2$ coherence times of NV$^-$ spins is a crucial issue for various quantum-device applications.  Although nitrogen ion implantation is a preferable route of fabricating NV$^-$, a sufficient amount of vacancies in the vicinity of an implanted N atom are required to create NV$^-$ centers with high efficiency.~\cite{Naydenov1,Pezzagna1,Schwartz} However, if vacancies accumulate excessively, paramagnetic multi-vacancy defects remain as residual radiation damage after annealing, and their magnetic fluctuations lead to decoherence of NV$^-$ spins. We observed coherence times, $T_2\sim2$~ms for NV$^-$ produced by 10~MeV micro-beam nitrogen ion implantation obtained in a 99.99\% $^{12}$C-enriched single crystal diamond by annealing at $1000^{\circ}$C. We also found that paramagnetic residual defects such as vacancy chains were significantly reduced by a modest increase of the annealing temperature from {\it{ensemble}} EPR measurements. We thus demonstrate that $^{12}$C enrichment together with selection of high-temperature annealing techniques can extend the $T_2$ of implanted NV$^-$ to much longer than 0.65~ms limited by the nuclear spin bath of natural abundance diamond. Our work opens new ways to build scalable quantum registers in a nuclear spin-free lattice by removing the obstacles that shorten $T_2$ in ion implanted diamond.

\begin{acknowledgments}
This study was carried out as {\textquoteleft}Strategic Japanese-German Joint Research Project' supported by JST and DFG (FOR1482, FOR1493 and SFB716), ERC, DARPA and the Alexander von Humboldt Foundation. We thank Brett C. Johnson for valuable discussions, and Kay D. Jahnke, Pascal Heller, Alexander Gerstmayr, and Andreas H\"au{\ss}ler for assistance with the experiments. 
\end{acknowledgments}

\end{document}